# Basic principles of postgrowth annealing of CdTe:Cl ingot to obtain semi-insulating crystals

**O. A. Matveev, A. I. Terent'ev**

*Ioffe Physicotechnical Institute, Russian Academy of Sciences, ul. Politekhnicheskaya 26, St. Petersburg, 194021 Russia*

- e-mail: Oleg.matveev@mail.ioffe.rssi.ru

The process of annealing of a CdTe:Cl ingot during its cooling after growth was studied. The annealing was performed in two stages: a high-temperature stage, with an approximate equality of chlorine and cadmium vacancy concentrations established at the thermodynamic equilibrium between the crystal and vapors of volatile components, and a low-temperature stage, with charged defects interacting to form neutral associations. The chlorine concentrations necessary to obtain semi-insulating crystals were determined for various ingot cooling rates in the high temperature stage. The dependence of the chlorine concentration $[Cl^{+}_{Te}]$ in the ingot on the temperature of annealing in the high-temperature stage was found. The carrier lifetimes and drift mobilities were obtained in relation to the temperature and cadmium vapor pressure in the postgrowth annealing of the ingot.

As is known, the low conductivity of CdTe crystals and the high free-carrier lifetimes and mobilities, needed for nuclear radiation detectors [1, 2], can be achieved in a chlorine-doped material owing to the self-compensation of charged atomic defects [3-6].

Previously [7, 8], we have studied the self-compensation in annealing of 3 x 3 x 12 mm$^3$ CdTe:Cl samples under controlled pressure of Cd and Te vapors, simulating the cooling of an ingot after crystal growth at $T \leq$ 980°C. The annealing produced semi-insulating samples with a conductivity of $\sigma = 10^{-10}$ $\Omega^{-1}$ cm$^{-1}$ and low free-carrier concentration $p(n) = (10^7$-$10^8)$ cm$^{-3}$. However, the mobility-lifetime products for electrons and holes in the annealed samples, $\mu_e\tau_e = 10^{-4}$ cm$^2$ V$^{-1}$, ($\mu_h\tau_h = 10^{-5}$ cm$^2$ V$^{-1}$, were much smaller than those in the as-grown ingot: $\mu_e\tau_e = 10^{-3}$ cm$^2$ V$^{-1}$, $\mu_h\tau_h = 10^{-4}$ cm$^2$ V$^{-1}$. Unfortunately, the annealing could not be performed in these studies at the highest temperatures, where post-growth annealing sets in, nor at low cadmium vapor pressure $P_{Cd} \longrightarrow P_{Cd}^{min}$. This was due to the fact that under these conditions the sublimation and transfer of the material into the cold part of an ampule caused changes not only at the surface, but also in the bulk of the sample, indicated by its nonuniform electrical conductivity.

In this paper, we report the results obtained by studying self-compensation in the annealing of an ingot immediately after its growth, with the annealing practically beginning with the compound crystallization temperature.

CdTe:Cl ingots weighing 0.5-1.0 kg are grown by horizontal planar crystallization under controlled cadmium vapor pressure [9]. During the growth, the material is doped with $N(Cl)$ of chlorine. During the post-growth cooling, the ingot is annealed, which enables the self-compensation of charged atomic crystal defects beginning with the highest temperatures. The self-compensation occurs in two annealing stages. In the first stage, there occurs high temperature annealing ($T_{ann}$ = 1070-800°C when the solubility of intrinsic atomic defects is high [10] and a high concentration of cadmium vacancies $[V^{-2}_{Cd}] + [V^{-1}_{Cd}] \geq [Cl^{+}_{Te}]$ can be obtained, which also should exceed the concentration of unintentional impurities. At this temperature, the association of donors and acceptors is weakly pronounced and can be neglected [3,10]. The second stage consists in low-temperature annealing ($T_{ann}$ = 800-400°C). This stage is dominated by the interaction of charged defects to form uncharged associations:

$(V^{-2}_{Cd}2Cl^{+}_{Te})^0, (V^{-1}_{Cd}2Cl^{+}_{Te})^0, (A^{-}D^{+})^0$. Primary defects, not bound into neutral associations, remain in the crystal: $V^{-2}_{Cd}$, $(V^{-2}_{Cd}Cl^{+}_{Te})^-$ and other defects, e.g., $(V^{-2}_{Cd} D^{+})^-$, $(A^{-2}Cl^{+}_{Te})^-$ (here $D$ and $A$ are the background donor and acceptor impurities), giving rise to energy levels in the band gap: $E_v + 0.9$ eV, $E_v + 0.14$ eV, and $E_v + (0.5$-$0.9)$ eV [11]. The carrier mobilities and lifetimes are determined by the concentrations of these crystal defects.

Hereinafter, no mention is made of the vacancy $V^{-1}_{Cd}$, present in much lesser amount compared with the $V^{-2}_{Cd}$

vacancy. The predominance of the $V^{-2}_{Cd}$ defect governs the self-compensation in CdTe [6,10].

Let us consider the self-compensation conditions in the high-temperature stage of the ingot annealing. When an ingot is kept at a constant temperature at thermodynamic equilibrium between CdTe and cadmium vapor with the pressure $P_{Cd}$ in the ampule for a long time $(t = 5\text{-}15\text{ h})$, the crystal composition corresponding to the vapor pressure and the related concentration of cadmium vacancies are established. The main condition for self-compensation is that VQJ be soluble at the annealing temperature. $T_{ann} = 800°C$ is the lowest temperature that allows a sufficiently high defect concentration $[V^{-2}_{Cd}] \geq [Cl^{+}_{Te}]$ to be obtained. At lower $T_{ann}$, the solubility of acceptor defects $V^{-2}_{Cd}$ at the crystal-gas equilibrium is not sufficiently high to enable the self-compensation of charged defects even at very low vapor pressures $P_{Cd} \approx P^{min}_{Cd}$.

Another condition for self-compensation is that the ratio of donor and acceptor concentrations $[V^{-2}_{Cd}] \geq [Cl^{+}_{Te}]$ be maintained constant during cooling at the high-temperature stage. It was found experimentally that the obtainment of a semi-insulating self-compensated material depends on the chlorine concentration. If the concentration of $[Cl^{+}_{Te}]$ defects exceeds the cadmium vacancy concentration during the ingot cooling, we obtain a high-conductivity material with $n \sim 10^{16}\text{cm}^{-3}$. The solubility of chlorine and cadmium vacancies decreases as the CdTe crystal temperature becomes lower [12, 13]. Consequently, two possibilities exist in ingot cooling. First, if the $[V^{-2}_{Cd}]$ and $[Cl^{+}_{Te}]$ concentrations decrease in approximately the same manner, the relation $[V^{-2}_{Cd}] \geq [Cl^{+}_{Te}]$ remains valid and a semi-insulating material is obtained. The second possibility is that the relation between the concentrations varies in such a way that $[Cl^{+}_{Te}]$ starts to exceed $[V^{-2}_{Cd}]$ and we obtain a high-conductivity material.

Either semi-insulating or high-conductivity CdTe can be obtained depending on $N(Cl)$ and the ingot cooling rate ($V_{CdTe}$) at the high-temperature stage (see Fig. 1). The $N(Cl)$ concentration includes chlorine dissolved in the CdTe ingot, i.e., electrically charged chlorine ($Cl^{+}_{Te}$), chlorine at grain boundaries, and also chlorine filling the gas space of the ampule and adsorbed by the graphite coating of the ampule and the container, etc. It seems reasonable that, the experimental conditions being the same, changes in $N(Cl)$ will lead to the corresponding changes in the chlorine content in the crystal itself. We used $N(Cl)$ concentrations no lower than $2 \times 10^{18}\text{ cm}^{-3}$, the minimal value necessary for crystal self-compensation to occur in CdTe growth from the melt [14], to $N(Cl) = 2 \times 10^{19}\text{ cm}^{-3}$, at which $CdCl_2$ precipitates at the grain boundaries [15]. To obtain semi-insulating crystals, the ingot cooling rate should be chosen so that the defect concentrations $[Cl^{+}_{Te}]$ and $[V^{-2}_{Cd}]$ are approximately the same during the entire high-temperature annealing stage. The curve in Fig. 1 divides its area into two regions corresponding to the two possible relations between the defect concentrations: $[V^{-2}_{Cd}] > [Cl^{+}_{Te}]$ and $[Cl^{+}_{Te}] > [V^{-2}_{Cd}]$ A semi-insulating material is obtained over the entire range of dopant concentrations $N(Cl) = 2 \times 10^{18}\text{-}2 \times 10^{19}\text{ cm}^{-3}$ when $V_{CdTe} = 2$ K/h (see Fig. 1). For higher ingot cooling rates, this region becomes narrower at the expense of high chlorine concentrations, when a high-conductivity material (region B in Fig. 1) is obtained. At the highest cooling rates $V_{CdTE} = 48$ K/h, a semi-insulating crystal is obtained only at the lowest concentrations $N(Cl) = (2\text{-}4) \times 10^{18}\text{ cm}^{-3}$.

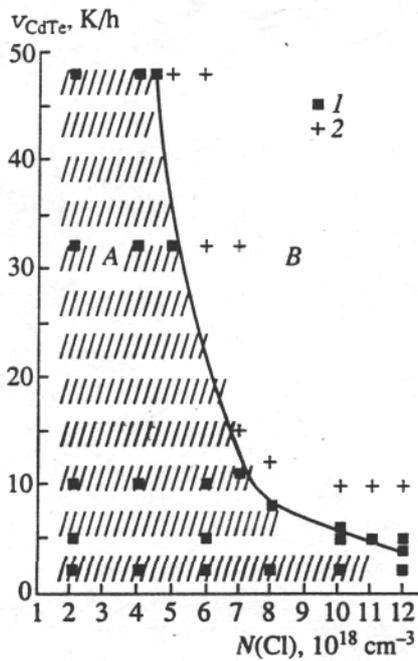

Fig. 1. Regions of low (A) and high (B) conductivities of CdTe:Cl crystals in the diagram showing the rate of ingot cooling (vorre) in the temperature range 1070-900°C against the chlorine concentration. Regions A and B correspond to concentrations: (A) $p \approx 10^8\text{ cm}^{-3}$ and (B) $n \approx 10^{16}\text{ cm}^{-3}$. Designations: (7) semi-insulating and (2) high-conductivity CdTe.

Under the annealing conditions represented by the high-conductivity region B (unshaded field in Fig. 1), high cooling rates at a high chlorine content in the ingot make it impossible to maintain equal concentrations of chlorine and cadmium vacancies [13,16]. This is associated with the fact that, in fast cooling, the $[Cl^{+}_{Te}]$ concentration has no time to attain the value corresponding to the equilibrium solubility of chlorine owing to its low diffusion coefficient, compared with that for the is equal to the atomic fraction of sinks $C$. In the case under consideration,

$$C \approx \frac{[Cl^+_{Te}]}{N_0} \approx \frac{[V^{-2}_{Cd}]}{N_0} \approx \frac{10^{16}\ cm^{-3}}{10^{22}\ cm^{-3}} = 10^{-6},$$

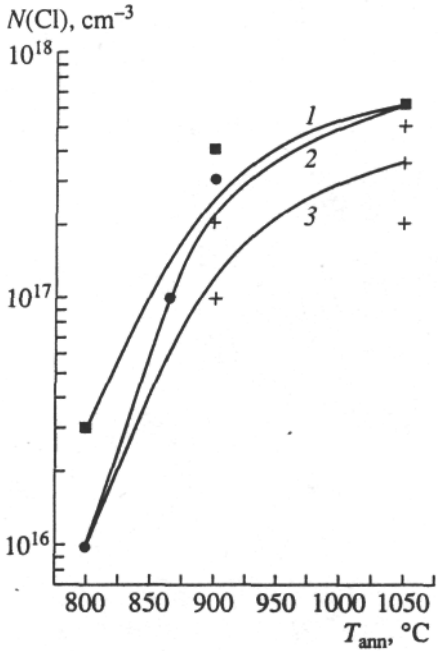

Fig. 2. Chlorine concentration $N(Cl)$ in the CdTe:Cl ingot, determined (7) by mass-spectral analysis, (2) by atomic sorption analysis, (5) from Hall data, vs. the annealing temperature $T_{ann}$.

cadmium vacancies. Consequently, the $Cl_{Te}$ impurity concentration in the crystal may exceed the concentration of $V_{Cd}$ vacancies, thus giving rise to a material with $n \sim 10^{16}\ cm^{-3}$.

Thus, the conditions for self-compensation in CdTe:Cl by high-temperature annealing of an ingot are determined for the entire high-temperature stage cooling after crystal growth in a wide range of doping levels and cooling rates.

The ingot annealing at the low-temperature stage is also performed satisfying the requirements that self-compensation should occur. In this stage, the ingot is cooled at an experimentally selected rate $v = 50$-$80$ K/h, which is sufficiently fast to prevent the attainment of a low cadmium vacancy concentration corresponding to these temperatures (800-400°C [17, 18]). Thus, the charged-defect concentration balance achieved in the high-temperature stage is not disturbed. At the same time, this cooling rate is quite sufficient to ensure the formation of associations of $Cl^+_{Te}$ donors with $V^{-2}_{Cd}$ acceptors, so that the crystal becomes "self-purified" [5]. The associations are formed as a result of defects migrating toward one another. The probability of association formation is evaluated using the theory of random walks [19]. The probability that the motion of a defect results in its arrival at a "sink" (association formation) nearly equals the probability that the lattice site at which the defect finds itself is a sink. This probability where $N_0$ is the atomic concentration per cubic centimeter. Therefore, the average number of hops necessary for arriving at a sink is approximately $ñ \approx 1/C = 10^6$. To determine the time necessary for association formation to be complete, we need to evaluate the time constant for the reduction in the concentration of defects τ. This constant can be calculated in terms of diffusion theory: $ñ = zv\tau$, where $z = 4$ is the coordination number and $v = D/a^2\gamma$ is the average frequency of jumps in the direction toward the sink, corresponding to the diffusion coefficient $D$ for a moving defect; $\gamma = 1$ for a cubic lattice; $a = 2.5 \times 10^{-8}$ cm is the interatomic distance in CdTe. Thus, we have $\tau = 1.6 \times 10^{-10}/D$. For the extreme temperatures in the second stage of annealing, 400 and 800°C, we have the following diffusion coefficients for the moving defects [13,16]:

$D(Cl) = 10^{-11}\ cm^2\ s^{-1}$, $D(V_{Cd}) = 10^{-10}\ cm^2\ s^{-1}$, and $D(Cl) = 10^{-9}\ cm^2\ s^{-1}$, $D(V_{Cd}) = 2 \times 10^{-8}\ cm^2\ s^{-1}$,

respectively. The cooling time of 6 h, chosen for our experiment, is $-10^3\tau$ even for the smallest diffusion coefficient $D$, and as a result, the process of defect association can be considered complete.

To obtain semi-insulating crystals, the relation $[Cl^+_{Te}] = [V^{-2}_{Cd}]$ should be fulfilled during both annealing stages. Therefore, it is important to know the chlorine concentration, and its temperature variation, directly in the crystal in order to specify in a substantiated way the necessary $[V^{-2}_{Cd}]$ concentration by maintaining the $P_{Cd}$ pressure.

The chlorine content in the ingot at varied annealing temperatures was determined by mass-spectrometric and atomic sorption analyses and, indirectly, by measuring the Hall constant of specially annealed single-crystal samples (Fig. 2). The free-carrier concentration corresponding to the chlorine (shallow donor) concentration, $n = [Cl^+_{Te}]$, was found from the Hall voltage [10]. The chlorine was transformed from bound states $(V^{-2}_{Cd}2Cl^+_{Te})^0$, $(A-Cl^+_{Te})^0$ into a donor state by annealing at a high pressure $P_{Cd}$ [7]. The results obtained by using both direct and indirect methods indicate that the chlorine concentration behaves in the same way in the CdTe ingot with the annealing temperature under postgrowth cooling conditions. The chlorine concentration determined from the Hall voltage is the lowest, compared with the values produced by other analytical methods, because it includes only electrically charged isolated $Cl^+_{Te}$ defects.

Figure 3 shows the carrier concentrations $n$ and p in relation to pressure $P_{Cd}$ for two annealing temperatures. For high $P_{Cd}$, $n$ reaches values of $10^{16}$ cm$^{-3}$, since the relation $[Cl^+_{Te}] > [V^{-2}_{Cd}]$ is valid under these conditions. The $Cl^+_{Te}$ defect gives rise to a shallow donor level in the band gap $E_c = 0.01$ eV, so its excess leads to a fast rise in $n$ in the crystal. The $V^{-2}_{Cd}$ defect and its associations give rise to deep levels at $E_v + (0.5-0.9)$ eV; therefore, $p$ also grows with decreasing $P_{Cd}$, but this process is much slower and $p$ increases only to $10^9$ cm$^{-3}$. The dependences of $n$ and $p$ on $P_{Cd}$, for annealing at 900 and 1070°C are surprisingly alike in shape at practically the same free-carrier concentrations. The minima in the dependences of $n$ and $p$ on pressure $P_{Cd}$ reliably characterize the maximum degree of self-compensation, $p(n)/[Cl^+_{Te}]$, when the donor and acceptor concentrations are equal, $[Cl_{Te}] = [V_{Cd}]$. The defect concentrations $[Cl^+_{Te}]$, $[V^{-2}_{Cd}]$ are not the same at different annealing temperatures in the high-temperature stage and after the association of the majority of defects in the low-temperature stage. This must manifest itself in carrier scattering in drift-mobility measurements and in carrier recombination in the measurement of lifetime.

Therefore, we studied the crystal parameters $\mu_e$, $\mu_h$, $\tau_e$, and $\tau_h$, the most sensitive to the extent to which the annealing process is complete, in relation to the ingot annealing temperature and $P_{Cd}$. These dependences were obtained for crystals with hole concentrations $p = (10^8 - 10^9)$ cm$^{-3}$ and with the largest $\mu\tau$ product for carriers, such crystals being characterized by a low content of unassociated $V^{-2}_{Cd}$ vacancies giving rise to a deep level in the band gap [11].

Figure 4 shows the drift mobility of electrons and holes as a function of the ingot annealing temperature. It can be seen that electrons and holes attain high mobility at relatively low annealing temperatures $T_{ann} = 800$-$900°C$; i.e., the mobility practically reaches its limiting value at the lower temperature in the interval corresponding to the high-temperature stage of ingot annealing. In semi-insulating CdTe:Cl, the carrier mobility is also determined by strains and inhomogeneities limiting the conducting part of the crystal or causing additional scattering [20]. At lower annealing temperatures, the solubility of charged atomic defects decreases. This diminishes the number of inhomogeneities and neutral associations of impurities and structural defects responsible for carrier scattering. The $\tau_e$ value also increases when the crystal annealing temperature is lowered (Fig. 5, curve 7), similarly to $\tau_{e,h}$, and presumably does so for the same reason. It can be seen from curve 2 in Fig. 5 that the hole lifetime grows with increasing crystal annealing temperature. Raising the temperature $T_{ann}$ leads to an increase in $V^{-2}_{Cd}$ solubility, both in absolute value and relative to that of the $V^{-1}_{Cd}$ vacancies [6,10]. The latter favors more complete association of these defects in the form of $(V^{-2}_{Cd}2Cl^+_{Te})^0$ [3].

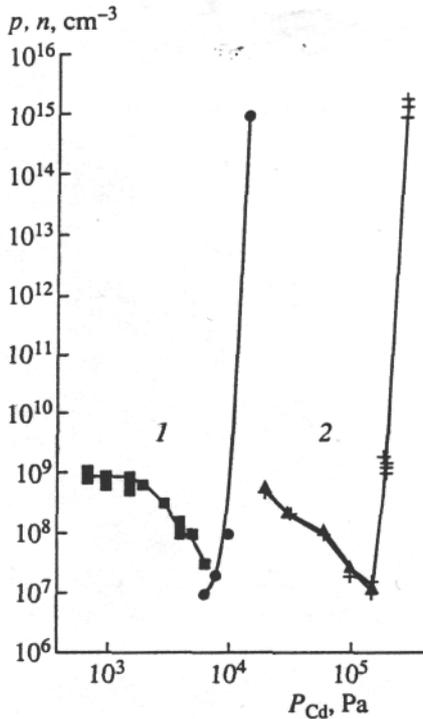

Fig. 3. Free carrier concentrations $(n, p)$ vs. the cadmium vapor pressure (Pea) for two temperatures $T_{ann}$ of CdTe:Cl ingot annealing: (*1*) 900 and (2) 1070°C.

Such a decrease in the content of $V^{-2}_{Cd}$ defects in the crystal leads to a lower concentration of deep levels in the band gap and, correspondingly, to a longer hole lifetime. Presumably, the decrease in the $[V^{-2}_{Cd}]$ concentration exerts a stronger influence on the hole lifetime $\tau_h$ than the dependence of the distribution of the defects determining $\tau_e$ on strains and inhomogeneities does.

Thus, attainment of the necessary $\tau_e$ and $\tau_h$ values requires some tradeoffs in the choice of ingot annealing temperatures in the high-temperature annealing stage. This criticality (see Fig. 5) demands a very careful choice of the ingot annealing temperatures, depending on specific requirements to the $\tau_e$ and $\tau_h$ values in the crystal, not only in the high-temperature stage, but also in the low-temperature annealing stage, when the $V^{-2}_{Cd}$ defects are eliminated via their association with donors. In connection with- this, it is worth studying the $\tau_e$, $\tau_h$, $\tau_e$, and $\tau_h$ dependences on the vacancy concentration determined by the vapor pressure $P_{Cd}$ in the ampule during ingot annealing. The maximum concentration of cadmium vacancies in the crystal is specified by the pressure $P_{Cd}$ —- $P_{Cd}^{min}$ With increasing $P_{Cd}$, this concentration decreases to values corresponding to *p*-type conduction (see Fig. 3, left-hand portions of curves *1* and *2*, for ingot annealing temperatures $T_{ann}$ = 900 and 1070°C, respectively). Figure 6 shows the dependences of $\tau_h$ and $\mu_h$ on $P_{Cd}$, at the same annealing temperatures. The tendency for these charge transport parameters of the crystal to decrease with increasing $P_{Cd}$ during annealing is also seen. This dependence of $\tau_h$ and $\mu_h$ on $P_{Cd}$ is explained as follows. The decrease in the $V^{-2}_{Cd}$ concentration in the ingot (with increasing $P_{Cd}$) in the high-temperature stage of annealing leads to a lower extent of $V^{-2}_{Cd}$ and $Cl^{+}_{Te}$ association in the low temperature stage. As a result, the $V^{-2}_{Cd}$ and $Cl^{+}_{Te}$ defects remaining unassociated after annealing contribute to a decrease in hole lifetime and mobility.

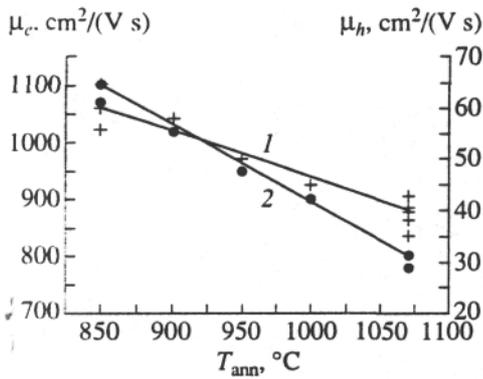

Fig. 4. Drift mobility of *(1)* electrons and (2) holes vs. the CdTe:Cl ingot annealing temperature $T_{ann}$

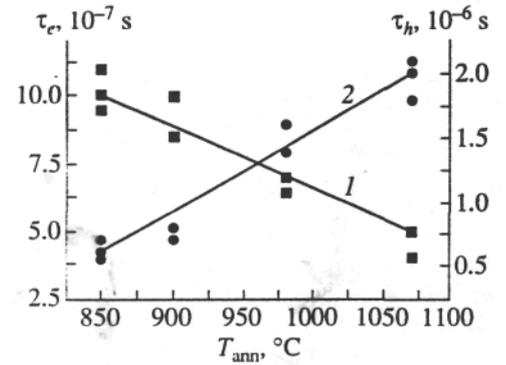

Fig.5 S. Lifetimes of electrons ($\tau_e$, 1 and holes ($\tau_h$, 2) vs. the CdTe:Cl ingot annealing temperature $T_{ann}$.

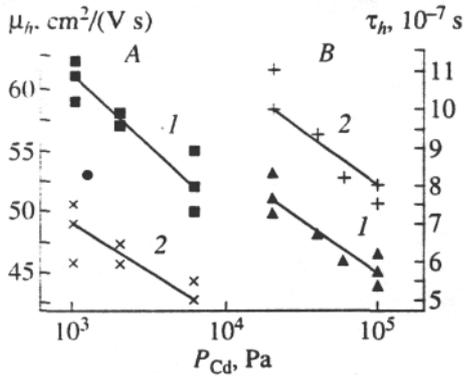

Fig. 6. (1) Drift mobility and (2) lifetime of holes vs. the cadmium vapor pressure $P_{Cd}$ in annealing of CdTe:Cl ingot at annealing temperatures $T_{ann}$, of *(A)* 900 and *(B)* 1070°C.

The influence of the ingot annealing temperature on the hole lifetime and mobility, illustrated in Figs. 4 and 5, can be also inferred from Fig. 6. For crystals annealed at $T_{ann}$ = 1070°C (interval *B of $P_{Cd}$* variation in Fig. 6) the lifetime is about 1.5 times longer than that for crystals annealed at $T_{ann}$=900°C (interval *A* in Fig. 6). For crystals annealed at $T_{ann}$=900°C, the hole mobility is about 1.2 times higher.

Thus, the annealing temperature is a key factor. Indeed, high $\tau_h$ values are attained only at high temperatures owing to the high solubility of $V^{-2}_{Cd}$. Raising the

$[V^{-2}_{Cd}]$ concentration by decreasing the pressure $P_{Cd}$ affects $\tau_h$, but to a smaller extent, precluding achievement of such high $\tau_h$ values. At low temperatures, high $\tau_h$, $\mu_e$, and $\mu_h$ are obtained owing to a decrease in the concentration of charged and neutral defects and the lower nonuniformity of their distribution.

No influence of $P_{Cd}$ on $\tau_e$ and $\mu_e$, was observed. This is explained by the fact that the changes in the $V^{-2}_{Cd}$ concentration, to which the quantities $\tau_h$ and $\mu_h$ for holes are sensitive, are insufficient for exerting a noticeable influence on $\tau_e$ and $\mu_e$.

Thus, we described here a procedure for two-stage postgrowth annealing of a CdTe:Cl ingot with programmed cooling to a low temperature (400°C). In the high-temperature stage of annealing, we studied the factors governing the relation between atomic defect concentrations $[Cl^{+}_{Te}] = [V^{-2}_{Cd}]$ This relation is the condition for sufficient self-compensation of charged defects in CdTe:Cl in the low-temperature annealing stage by way of defect association into neutral complexes. The rates of ingot cooling corresponding to different levels of chlorine doping were established, ensuring self-compensation. We determined the manner in which the chlorine solubility decreases as temperature becomes lower in the course of ingot cooling. In the process, exact self-compensation and, conse-quently, large $\mu\tau$ products for carriers in the crystal are achieved by adjusting the concentration $[V^{-2}_{Cd}] \geq 0.5[Cl^{+}_{Te}]$. A tendency was observed for $\mu_e$, $\mu_h$ and $\tau_e$ to increase as the ingot annealing temperature is lowered, which is explained by a decrease in the content of charged defects in the crystal at low temperature. An opposite tendency was observed for the hole lifetime, $\tau_h$, which becomes longer with increasing annealing temperature. This may be due to a decrease hi the vacancy concentration $[V^{-2}_{Cd}]$ in the process of their association into neutral complexes $(V^{-2}_{Cd}2Cl^{+}_{Te})^0$, occurring to a greater extent during high-temperature annealing.

To conclude, the proposed two-stage postgrowth annealing of the ingot allows extremely precise control over the self-compensation and "self-purification" processes and enables growth of CdTe:Cl semi-insulating crystals with good transport properties. We believe that an analogous two-stage annealing will make it possible to produce other II-VI compounds and their solid solutions with desirable characteristics by horizontal planar crystallization, since in these materials the self-compensation mechanism is known to be clearly pronounced, similarly to CdTe crystals.

ACKNOWLEDGMENTS

This study was supported by INTAS, grant no. 99-01456.